\documentclass[a4paper,11pt]{article}
\usepackage{pos}
\usepackage{bigstrut}

\usepackage[normalem]{ulem}
\usepackage{soul}

\title{LASS, the numerics}
\ShortTitle{LASS, the numerics}

\author*[a,b]{A.~Kardos}
\author[c]{G.~Bevilacqua}
\author[d]{B.~Chargeishvili}
\author[d]{S.~O.~Moch}
\author[b,a]{Z.~Trocsanyi}

\affiliation[a]{Department of Experimental Physics, Institute of Physics, 
                Faculty of Science and Technology, University of Debrecen,\\
                4010 Debrecen, PO Box 105, Hungary
}

\affiliation[b]{Institute for Theoretical Physics, ELTE E\"otv\"os Lor\'and University,\\
                P\'azm\'any P\'eter 1/A, H--1117 Budapest, Hungary
}

\affiliation[c]{Institute of Nuclear and Particle Physics, NCSR "Demokritos",\\
                15341 Agia Paraskevi, Greece}

\affiliation[d]{II. Institut f\"ur Theoretische Physik, Universit\"at Hamburg,\\
                Luruper Chaussee 149, 22761 Hamburg, Germany}

\emailAdd{kardos.adam@science.unideb.hu}
\emailAdd{bevilacqua@inp.demokritos.gr}
\emailAdd{bakar.chargeishvili@desy.de}
\emailAdd{sven-olaf.moch@desy.de}
\emailAdd{zoltan.trocsanyi@cern.ch}

\abstract{
We summarize our efforts to create a numerical implementation of the Local Analytical Subtraction Scheme (LASS) for obtaining NNLO QCD predictions in electron-positron collisions. We focus on the regularization of double-real radiation contributions, which are typically the most computationally intensive part of NNLO calculations. We describe the structure of the subtraction terms in LASS and our approach to numerically validate them. Using arbitrary precision arithmetic, we demonstrate the proper convergence of individual subtraction terms, overlap removal terms, and spurious singularity cancellations in various singular limits. We show results for the specific case of three-jet production in e+e- collisions. 
Our results validate the correctness of both the LASS formalism and our numerical implementation, setting the stage for full NNLO calculations using this method.
}

\FullConference{Loops and Legs in Quantum Field Theory (LL2024)\\
 14-19, April, 2024\\
Wittenberg, Germany\\}

\newcommand{\fig}[1]{Fig.~\ref{#1} }
\newcommand{\aS}{\ensuremath{\alpha_{\rm S}} }
\newcommand{\ud}{\ensuremath{{\rm d}\,}}
\newcommand{\lo}{\ensuremath{{\rm LO}} }
\newcommand{\nlo}{\ensuremath{{\rm NLO}} }
\newcommand{\nnlo}{\ensuremath{{\rm NNLO}} }
\newcommand{\LassSoft}[2]{ \ensuremath{\overline{\mathbf S}_{#1}{#2}} }
\newcommand{\LassColl}[3]{ \ensuremath{\overline{\mathbf C}_{#1}^{#2}{#3}} }
\newcommand{\LassSC}[3]{ \ensuremath{\overline{\mathbf{SC}}_{#1}^{#2}{#3}} }
\newcommand{\LassHColl}[3]{ \ensuremath{\overline{\mathbf{HC}}_{#1}^{#2}{#3}} }
\newcommand{\LassSHC}[3]{ \ensuremath{\overline{\mathbf{SHC}}_{#1}^{#2}{#3}} }

\newcommand{\zsecfunc}[1]{ \ensuremath{\mathcal{Z}_{#1}} }
\newcommand{\zbsecfunc}[1]{ \ensuremath{\overline{\mathcal{Z}}_{#1}} }
\newcommand{\LassSubs}[1]{ \ensuremath{K^{(#1)}} }
\newcommand{\qq}{ \ensuremath{{\rm q}} }
\newcommand{\qaq}{ \ensuremath{\bar{\rm q}} }
\newcommand{\qr}{ \ensuremath{{\rm r}} }
\newcommand{\qar}{ \ensuremath{\bar{\rm r}} }
\newcommand{\mpfun}{\texttt{MPFUN2020}}
\begin{document}
\maketitle

\section{Introduction}
%{{{
We are living in the precision era of LHC physics: precise
measurements are common-place today.  For example, we refer to the
$p_\bot$ distribution for lepton-pairs produced in the Drell-Yan
process at low-$p_\bot$ values \cite{CMS:2018mdl,ATLAS:2019zci}.
High-precision measurements are necessary but not sufficient to drive
the physics potential of the LHC to the maximum. In order to harvest the
most information equally high-precision predictions are needed from
the theory side. Such predictions are needed to successful searches for
physics beyond of the standard model as well as provide the key to
precision phenomenology.

For high-precision predictions higher-order calculations are needed in
both sectors of the standard model: in quantum chromodynamics (QCD)
and the electroweak sector. These calculations are cumbersome,
technologically challenging and CPU-time hungry.  High-accuracy
measurements impose yet one more requirement on these calculations,
namely not only the formal, but also the numerical accuracy of the 
computations should be decent enough to meet the expected precision of
the predictions.

The QCD corrections are usually sizable at the LHC, which necessitates
to go beyond the next-to-leading order (NLO) in the computations.
Hence several calculational schemes were constructed for making QCD
predictions at the  next-to-next-to-leading order (NNLO) accuracy. Although these methods can be
used to produce cross section values, the uncertainty of the
required numerical integrations is often insufficient for the expected
precision. The trivial solution to this problem is the increase of
computational time, which however comes with a high price. The CPU
time devoted to such calculations has reached millions of hours. 
As science is a part of society, it cannot be insensitive to problems
of society. Energy efficiency is paramount for a sustainable society,
which imposes an extra constraint on our calculation schemes. While
high theoretical accuracy and numerical precision are expected,
those should be reached with as little CPU time as possible. This
requirement leads us to investigate any new idea in regards to
calculating higher-order corrections promising higher energy efficiency.

One such idea is the Local Analytical Subtraction Scheme (LASS)
\cite{Magnea:2018hab,Magnea:2020trj,Bertolotti:2022aih}. 
While the LASS has been developed primarily for electron-positron
collisions, the simplicity of its subtraction structure and the
resulting integrated terms suggest possible generalization to collisions 
with initial states involving QCD partons. Despite the theoretical
foundation laid out in the original publications, a comprehensive
numerical implementation at NNLO remains lacking. Our objective is
to address this gap by developing a numerical implementation of the
LASS formalism. In the following sections, we outline our efforts to
implement and validate the essential ingredients required for the $n+2$
parton contribution, i.e.~the regularized double-real radiation
contribution, within the framework of an NNLO QCD calculation.

%}}} Introduction

\section{LASS from the practitioner's point of view}
%{{{
In perturbative QCD the cross section is defined as a series in the
strong coupling \aS. The fully differential cross section can formally
be written as:
\begin{align}
  \ud\sigma 
  &=
  \ud\sigma_{\lo}
  +
  \ud\sigma_{\nlo}
  +
  \ud\sigma_{\nnlo}
  +
  \dots
  \,,
\end{align}
where we labelled the first, non-vanishing contribution as LO (for
leading order), and kept the next two orders (NLO and NNLO).
The LO contribution is the fully differential Born cross section,
\begin{equation}
\ud\sigma_{\lo} = \ud\sigma^{\rm B}
\,.
\end{equation}

Beyond LO several different ingredients build up the cross section
contribution. These ingredients are categorized according to the number
of additional partons present as virtual particles in loops, or real
ones in the final state. The correction at NLO is a sum of two terms,
the single real and virtual contributions, while at NNLO there are
double-real, real-virtual and double-virtual corrections,
\begin{align}
  \,\quad
  \ud\sigma_{\nlo}
  =
  \ud\sigma^{\rm R}
  +
  \ud\sigma^{\rm V}
  \,,\quad
  \ud\sigma_{\nnlo}
  =
  \ud\sigma^{\rm RR}
  +
  \ud\sigma^{\rm RV}
  +
  \ud\sigma^{\rm VV}
  \,.
\end{align}

In this contribution we are going to focus on the regularized real-emission
contributions being the most CPU time consuming elements of N$^n$LO QCD
calculations. We shall discuss the case of leptons in the initial state
as the LASS is published for electron-positron collisions. We shall use
the definitions of Ref.~\cite{Bertolotti:2022aih}, in particular
introduce the concept of scaleless energies and angle variables
characterizing softness and collinearity of parton emissions in terms
of two-particle invariants $s_{ij} = 2p_i\cdot p_j$,
\begin{align}
  e_i
  &=
  \frac{s_{iq}}{s}
  \,,
  \quad
  w_{ij} 
  =
  \frac{s \, s_{ij}}{s_{iq} s_{jq}}
  =
  \frac{s_{ij}}{s}
    \frac{s^2}{s_{iq} s_{jq}}
  \,,
\end{align}
where $q$ is the total incoming four-momentum, $q^{\mu} = (\sqrt{s},
0,0,0)$. The same quantities can also be expressed through energies and
angles as
\begin{align}
  e_i 
  &=
  \frac{2 E_i}{\sqrt{s}}
  \,,
  \quad
  w_{ij}
  =
  \frac{1}{2}\left(1 - \cos\theta_{ij}\right)
  \,.
\end{align}
These variables give a convenient way to parametrize the various
kinematic singularities at NLO and to define appropriate subtraction
terms $\LassSoft{i}{R}$ in the single-soft ($e_i\to 0$) and
$\LassColl{ij}{}{R}$ in the single-collinear ($w_{ij}\to 0$) limits.
In these subtraction terms an operator-like behavior is understood in
order to take into account color- (soft) and spin-correlations
(collinear) and to replace real-emission dynamics with underlying,
non-singular Born multiplied by a factor  mimicking the singular behavior
of original real radiation squared matrix element (SME)
$R = |\mathcal{M}_{n+1}|^2$. The soft
and collinear regions are not disjoint. The overlaps should be removed
in order to avoid double subtractions. Considering all kinematic
singularities the regularized real emission can be cast into a form of
\begin{equation}
  R -
  \sum_{i}
  \left[
    \LassSoft{i}{{}{}} + \sum_{j>i} \LassColl{ij}{}{{}}
    \left( 1 - \LassSoft{i}{{}{}} - \LassSoft{j}{{}{}} \right)
  \right] R
  \,,
\label{eq:single}
\end{equation}
where we used the operator-like actions in the subtraction terms. 
Because the overlap removal can be achieved by dropping soft terms from
two-particle Altarelli-Parisi kernels the regularized real-radiation
SME can be greatly simplified and written in the short form $R - \LassSubs{1}$
where
\begin{equation}
  \LassSubs{1} =
  \sum_{i} \LassSoft{i}{{}{}} R + \sum_{i} \sum_{j>i} \LassHColl{ij}{}{{}} R
  \,.
\label{eq:K1}
\end{equation}
The new notation for the modified collinear subtractions stands
for the hard-collinear terms with the meaning of removed soft pieces from 
two-particle Altarelli-Parisi kernels.

At NNLO the set of possible kinematic singularities is enlarged,
but it is still possible to characterize and define subtractions
using only the scaleless variables previously defined. The single
unresolved subtractions are analogous to those in Eq.~\eqref{eq:single}
with $R$ replaced with $RR = |\mathcal{M}_{n+2}|^2$. The double
soft ($e_i,\,e_j\to 0$) and soft-collinear singularities are
regularized as
\begin{equation}
  RR - \LassSoft{ij}{RR}
  \,,\quad
  RR - \LassSC{ijk}{}{RR}
\,,
\end{equation}
while the triple ($w_{ij},\,w_{ik},\,w_{jk}\to 0$) and double collinear
($w_{ij},\,w_{kl}\to 0$) singularities are regularized as
\begin{equation}
  RR - \LassColl{ijk}{}{RR}
  \,,\quad
  RR - \LassColl{ijkl}{}{RR}
  \,.
\end{equation}
Overlapping singular regions are also present among subtraction terms
for double unresolved limits. To remove these overlaps, in addition to
the single hard-collinear limit of Eq.~\eqref{eq:K1} with the replacement
$R\to RR$, we introduce the following combinations, defining the
hard-collinear limits:
\begin{align}
  \LassHColl{ijk}{}{RR} &=
  \LassColl{ijk}{}{}
  \left( 1 - \LassSoft{ij}{} - \LassSoft{ik}{} - \LassSoft{jk}{} \right) RR
  \,,\nonumber\\
  \LassSHC{ijk}{}{} \left( 1 - \LassColl{ijk}{}{} \right) RR &=
  \LassSC{ijk}{}{}
  \left( 1 - \LassSoft{ij}{} - \LassSoft{ik}{} \right)
  \left( 1 - \LassColl{ijk}{}{} \right) RR
  \,,\nonumber\\
  \LassHColl{ijkl}{}{RR} &=
  \LassColl{ijkl}{}{}
  \left(
    1 + \LassSoft{ik}{} + \LassSoft{jk}{} + \LassSoft{il}{} + \LassSoft{jl}{}
    - \LassSC{ikl}{}{} - \LassSC{jkl}{}{} - \LassSC{kij}{}{} - \LassSC{lij}{}{}
  \right) RR
  \,.
\end{align}
Then the subtraction terms for the regularization of singularities in the
double unresolved regions for the double-real radiation SME,
$RR - \LassSubs{1} - \LassSubs{2}$, can be grouped using the
hard-collinear notation as follows:
\begin{align}
  \LassSubs{1} &= \sum_{i} \LassSoft{i}{{}{RR}} +
  \sum_{i} \sum_{j>i} \LassHColl{ij}{}{{RR}}
  \,,\\
  \LassSubs{2} &= \sum_{i} \sum_{j>i} \LassSoft{ij}{}{{RR}}
  + \sum_{i} \sum_{j>i} \sum_{k>j} \LassHColl{ijk}{}{RR}
  + \nonumber\\ &+
  \sum_{i} \sum_{j\ne i} \sum_{\substack{k>j\\k\ne i}} \LassSHC{ijk}{}{}
  \left( 1 - \LassColl{ijk}{}{} \right) RR +
  \sum_{i} \sum_{j > i} \sum_{\substack{k>i\\ k\ne j}}
  \sum_{\substack{l>k\\ l\ne j}} \LassHColl{ijkl}{}{RR}
  \,.
\end{align}
This construction allows for a {\em partial} kinematic regularization
expressed in the following symbolic form:
\begin{align}
  \left. \left( RR - \LassSubs{1} \right)\right|_{\rm singly-unres.}
  &= {\rm finite}
  %\,,\\
  \,,\quad
  \left. \left( RR - \LassSubs{2} \right)\right|_{\rm doubly-unres.}
  %&
  = {\rm finite}
  \,.
\end{align}
However,
as a consequence of \LassSubs{1} developing spurious singularities in
doubly-unresolved and \LassSubs{2} developing spurious singularities in
singly-unresolved regions of phase space,
\begin{align}
  \left.
  \left( RR - \LassSubs{1} - \LassSubs{2} \right)
  \right|_{\rm singly-\,\&\,doubly-unres.}
  &\ne {\rm finite}
  \,.
\end{align}
To have full kinematic regularization an additional collection of terms
has to be introduced:
\begin{align}
  \left.
  \left(
    RR - \LassSubs{1} - \LassSubs{2} - \LassSubs{12} \right)
  \right|_{\rm singly-\,\&\,doubly-unres.}
  = {\rm finite}
  \,,
\end{align}
where the role of the newly introduced contribution, \LassSubs{12} is
two-fold. It regularizes spurious singularities of \LassSubs{1} in doubly-
and those of \LassSubs{2} in singly-unresolved regions.

In this contribution we focus on the regularization of the double-real
radiation.  In order not to change the complete NNLO correction, the
subtraction terms $K^{(i)}$ ($i = 1$, 2, or 12) will be integrated
analytically over the one- or two-particle radiation phase spaces
depending on the subtraction terms, and subsequently added to the
real-virtual and double virtual contributions with reduced particle
multiplicity in final-state. A companion contribution
\cite{Chargeishvili:2024LL24} gives details how this is done in the case
of the correction with $n+1$ partons in the final state.

Due to its double role, the definition of the \LassSubs{12} subtraction
is fairly delicate, which is facilitated by the introduction of
sectors, where only a subset of the singular regions exists.
The sectors are specified by introducing corresponding sector functions
\begin{align}
  \zsecfunc{ijk}
  \,,\quad
  \zsecfunc{ijkl}
  \,,
\end{align}
whose detailed construction is presented in Sections 3.2 and 3.6 of
Ref.~\cite{Bertolotti:2022aih}.  In order not to change the physical
cross section, the sector functions must fulfil the important sum rule
\begin{align}
  1 &=
  \sum_{i<j<k} \zsecfunc{ijk}
+ \sum_{\substack{i\,,j\\j>i}}
  \sum_{\substack{k>i\\k\ne j}}
  \sum_{\substack{l>k\\l\ne j}}
  \zsecfunc{ijkl}
  \,,
\end{align}
which offers a way to continuously partition the double-real emission
phase space:
\begin{align}
\mathrm{d}\Phi_{n+2}
  &=
  \sum_{i<j<k} \zsecfunc{ijk} \mathrm{d}\Phi_{n+2}
+ \sum_{\substack{i\,,j\\j>i}}
  \sum_{\substack{k>i\\k\ne j}}
  \sum_{\substack{l>k\\l\ne j}}
  \zsecfunc{ijkl} \mathrm{d}\Phi_{n+2}
  \,.
\end{align}
The sectors defined by sector function $\zsecfunc{ijk}$ contain singular
regions where the subtraction terms 
\begin{equation}
    \left(
      \LassSoft{i}{}
      \,,
      \LassSoft{j}{}
      \,,
      \LassSoft{k}{}
      \,,
      \LassColl{ij}{}{}
      \,,
      \LassColl{ik}{}{}
      \,,
      \LassColl{jk}{}{}
      \,,\bigstrut\\
      \LassSoft{ij}{}
      \,,
      \LassSoft{ik}{}
      \,,
      \LassSoft{jk}{}
      \,,
      \LassSoft{i}{}\LassColl{jk}{}{}
      \,,
      \LassSoft{j}{}\LassColl{ik}{}{}
      \,,
      \LassSoft{k}{}\LassColl{ij}{}{}
      \,,
      \LassColl{ijk}{}{}
  \right)RR
\end{equation}
becomes singular, while the sectors corresponding to $\zsecfunc{ijkl}$
contain singular regions where 
\begin{equation}
  \left(
      \LassSoft{i}{}
      \,,
      \LassSoft{j}{}
      \,,
      \LassSoft{k}{}
      \,,
      \LassSoft{l}{}
      \,,
      \LassColl{ij}{}{}
      \,,
      \LassColl{kl}{}{}
      \,,
      \LassSoft{ik}{}
      \,,
      \LassSoft{il}{}
      \,,
      \LassSoft{jk}{}
      \,,
      \LassSoft{jl}{}
      \,,
      \LassSoft{i}{}\LassColl{kl}{}{}
      \,,
      \LassSoft{j}{}\LassColl{kl}{}{}
      \,,
      \LassSoft{k}{}\LassColl{ij}{}{}
      \,,
      \LassSoft{l}{}\LassColl{ij}{}{}
      \,,
      \LassColl{ijkl}{}{}
      \right)RR
\end{equation}
become singular. The operators are meant to affect not just the
dynamics but also these sector functions.  The limit(s) corresponding
to the subtraction term are strictly taken in the sector function
resulting in a factorized form as defined in Eqs.~(3.12) to (3.15) of
Ref.~\cite{Magnea:2018hab}.  There is ample freedom choosing the sector
functions permitted as long as the analytic integration of the
subtractions is possible.

A subtraction term can contribute in many sectors. The collinear terms
necessitate the introduction of reference momenta to define Sudakov
parameters. Iterated Catani-Seymour mappings \cite{Catani:1996vz} are
used to factorize the real-radiation phase space into a phase space with
lower multiplicity times a one- or two-body phase space factor. These
mappings depend on the sector and the subtraction at hand. To have a
working subtraction method the integration over factorized one- or
two-body phase spaces should be performed analytically, which is 
possible if the action of the soft and hard-collinear operators on
the sector functions is to collapse them to unity or factorize them
with mapped momenta (lower multiplicity) to allow for analytical
integrations.

To demonstrate explicitly how these actions take place, consider the
definite example of a single-collinear subtraction term $\LassColl{ij}{}{RR}$.
This subtraction becomes singular in many sectors, namely in
\begin{align}
  \left\{\zsecfunc{ijk}\,:\,k\ne i\,,j\right\}
  \bigcup
  \left\{\zsecfunc{ijkl}\,:\,k\,,l\ne i\,,j\,,k<l\right\}
  \,.
\end{align}
Thus, the collinear subtractions for the $ij$ splitting pair can be
written as: 
\begin{align}
\sum_{k\ne i,j}
\bigg[ \LassColl{ij}{}{RR} \, \zsecfunc{ijk}
+ \sum_{\substack{l>k\\ k,l\ne i,j}} \LassColl{ij}{}{RR} \, \zsecfunc{ijkl}
\bigg]
=
\sum_{k\ne \bar{\jmath}}
\bigg[ \zbsecfunc{\bar{\jmath}k} \LassColl{ij}{}{RR}
+ \sum_{\substack{l>k\\ k,l\ne \bar{\jmath}} }\zbsecfunc{kl} \LassColl{ij}{}{RR}
\bigg]
\,,
\end{align}
where the bar over the sector function signifies that it is calculated
with mapped momenta and $\bar{\jmath}$ stands for the mother parton
(with mapped momentum) for the $ij$ splitting pair. The reference
momentum can change from sector to sector, hence the subtraction term
cannot be factorized. Still, analytical integration over one-particle
radiation phase space is possible with sector functions depending
solely on the momenta of the underlying kinematics.

In a similar fashion we can consider the case of a double-soft
subtraction term contributing to both three- and four-index sectors
\begin{align}
  \left\{\zsecfunc{ijk}\,:\,k\ne i\,,j\right\}
  \bigcup
  \left\{\zsecfunc{ikjl}\,:\,k\ne i\,,j\,,l\ne i\,,j\,,k\right\}
  \,.
\end{align}
Looking at the double-soft operator acting on the sector functions the
following limiting behavior is observed (cf.~Eqs.~(3.4), (C.94) and
(C.95) of Ref.~\cite{Bertolotti:2022aih}):
\begin{align}
  \LassSoft{ij}{}\zsecfunc{ijk}
  &=
  \frac{
    \sigma_{ijjk} + \sigma_{ikjk} + (i\leftrightarrow j)
  }
  {
    \sum_{m\ne i}\sum_{n\ne i,j}\sigma_{imjn} + (i\leftrightarrow j)
  }
  \,,\quad
  \LassSoft{ij}{}\zsecfunc{ikjl}
  =
  \frac{
    \sigma_{ikjl} + \sigma_{jlik}
  }
  {
    \sum_{m\ne i}\sum_{n\ne i,j}\sigma_{imjn} + (i\leftrightarrow j)
  }
  \,.
\end{align}
Then the full double-soft subtraction for the $ij$ soft pair can be rewritten as
\begin{align}
  \sum_{k\ne i,j}
  \left[
    \LassSoft{ij}{RR}
      \,
      \zsecfunc{ijk}
    +
    \sum_{l\ne i,j,k}
    \LassSoft{ij}{RR}
      \,
      \zsecfunc{ikjl}
  \right]
  =
  \LassSoft{ij}{RR}
  \left[
    \LassSoft{ij}{}
      \,
      \zsecfunc{ijk}
    +
    \sum_{l\ne i,j,k}
    \LassSoft{ij}{}
      \,
      \zsecfunc{ikjl}
  \right]
  =
  \LassSoft{ij}{RR}
  \,,
\end{align}
where factorization was made possible by the subtraction being independent
of any reference momentum, hence sector functions. The limit sector 
function in the square bracket, by construction, also obey a sum rule, thus
adding up to one leaving only a sector-independent subtraction term.

%}}} LASS from the practitioner's point of view

\section{Numerical checks}
%{{{
As discussed, the construction of the complete LASS scheme requires many
steps:
\begin{itemize}
\itemsep=-2pt
  \item mappings to underlying kinematics,
  \item reference momentum choices,
  \item Sudakov parametrizations,
  \item spin- and/or color-correlated SME of reduced kinematics,
  \item overlap removal.
\end{itemize}

The construction of a complete partonic Monte Carlo program requires
careful numerical checks. To this end we selected the specific process
of production of three hadronic jets in electron-positron annihilation.%
\footnote{The LASS scheme uses iterated Catani-Seymour mappings with
reference (recoil) momentum chosen from the final state, thus it is not
suitable for two-jet production in electron-positron annihilation.}
Three-jet production at NNLO in QCD has three different classes of subprocesses with
the highest multiplicity in the final state:
\begin{align}
  e^+\,e^-\to \qq\qaq\qr\qar g
  \,,\quad
  e^+\,e^-\to \qq\qaq\qq\qaq g
  \,,\quad
  e^+\,e^-\to \qq\qaq ggg
  \,.
\end{align}
In order to numerically test the subtraction scheme we created a
\texttt{Fortran~90} program which uses the \mpfun\ package
\cite{Bailey:mpfun} to evaluate both dynamics and subtractions in
arbitrary precision%
\footnote{For our tests it was sufficient to use 50 working digits.}.
In order to keep numerical precision consistent we re-implemented the tree-level SMEs describing $e^+\,e^- \to n$ partons ($n \le 5$) using appropriate data types.
We used the same framework to test both
the regularized double-real and real-virtual \cite{Chargeishvili:2024LL24}
contributions.  The latter involves one-loop contributions up to four
partons in the final state, so we also re-implemented these using data
types and special functions provided by the \mpfun\ package. The
limiting behavior affects Sudakov parameters that appear in the arguments
of classical polylogarithms, whose function instances were changed to
special variants supporting arbitrary precision.

To perform complete tests of the subtraction terms, all possible singular limits have to be iteratively approached. This is achieved by generating an underlying Born phase space point (also in arbitrary precision) and sequentially increasing the multiplicity of the final state applying the
inverted Catani-Seymour mapping.

To characterize the energies (softness)
and enclosed angle (collinearity) of the produced pairs a
$\lambda\in[0,1]$ parameter was defined and used. From iteration to
iteration appropriately decreasing $\lambda$ by (half) an order of
magnitude\footnote{Whenever a soft-collinear limit was probed $\lambda$
was used to characterize energy of the soft-candidate parton,
which was decreased by an order of magnitude, while the square-root of
$\lambda$ was used to characterize collinearity of the collinear pair
to maintain proper scaling between these limits.} all possible NNLO
limits could be approached.

The code makes it possible to validate subtraction terms at multiple levels:
\begin{itemize}
\itemsep=-2pt
\item Individual subtraction terms compared to the appropriate
radiation SME in their restrictive limits.
\item Individual overlap terms compared to a single subtraction checking
proper overlap removal.
\item Individual terms defined to cancel spurious singularities compared
to a term from $K^{(1)}$ or $K^{(2)}$ to check cancellation of the
spurious singularity.
\item Full set of subtractions compared to the radiation SME in all
possible physical limits.
\end{itemize}
In the following we show examples for each of these cases demonstrating
the capabilities of our code and of the subtraction scheme. To be
definite we selected the most complicated subprocess with three gluons
in the final state. To refer to various limits, we use the labelling
$e^+_{1}\,e^-_{2}\to \qq_{3}\,\qaq_{4}\,g_{5}\,g_{6}\,g_{7}$.

\begin{figure}
\centering
\includegraphics[width=0.49\textwidth]{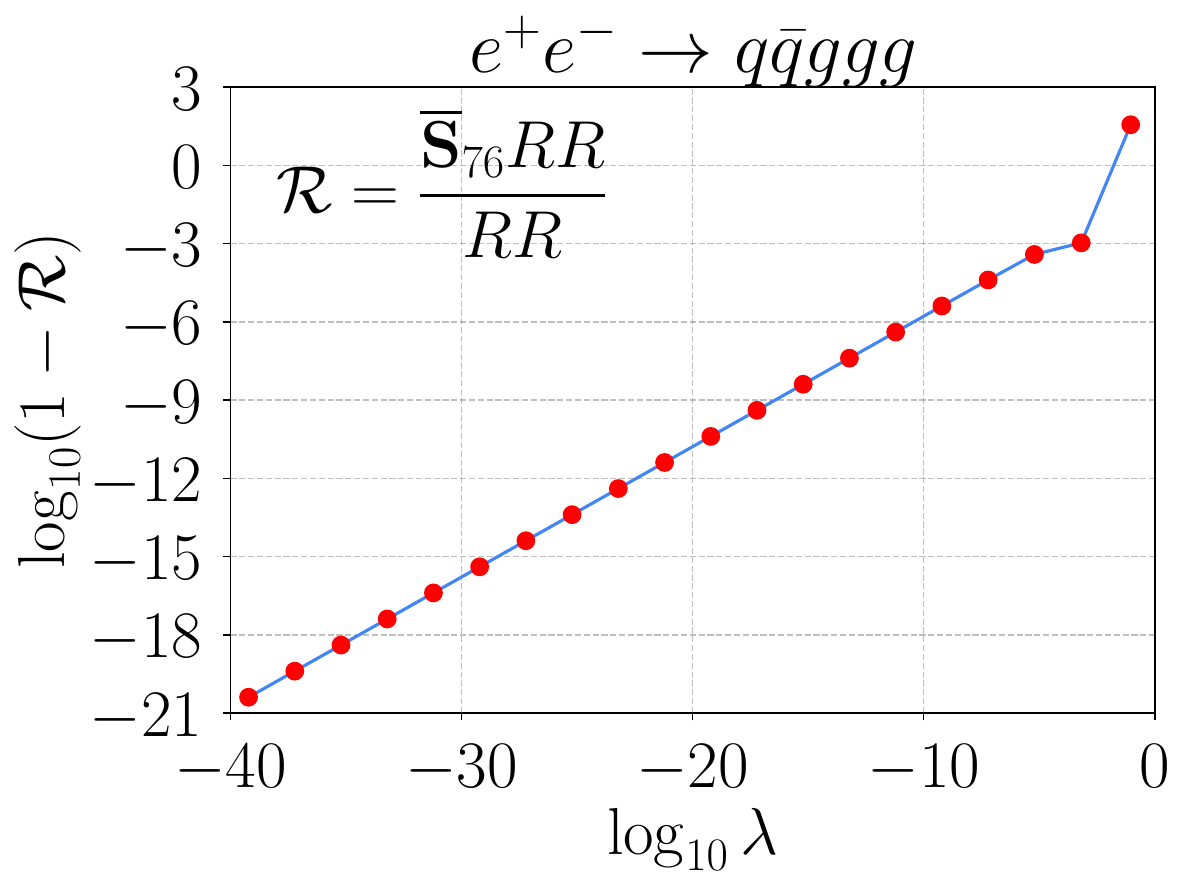}
~
\includegraphics[width=0.49\textwidth]{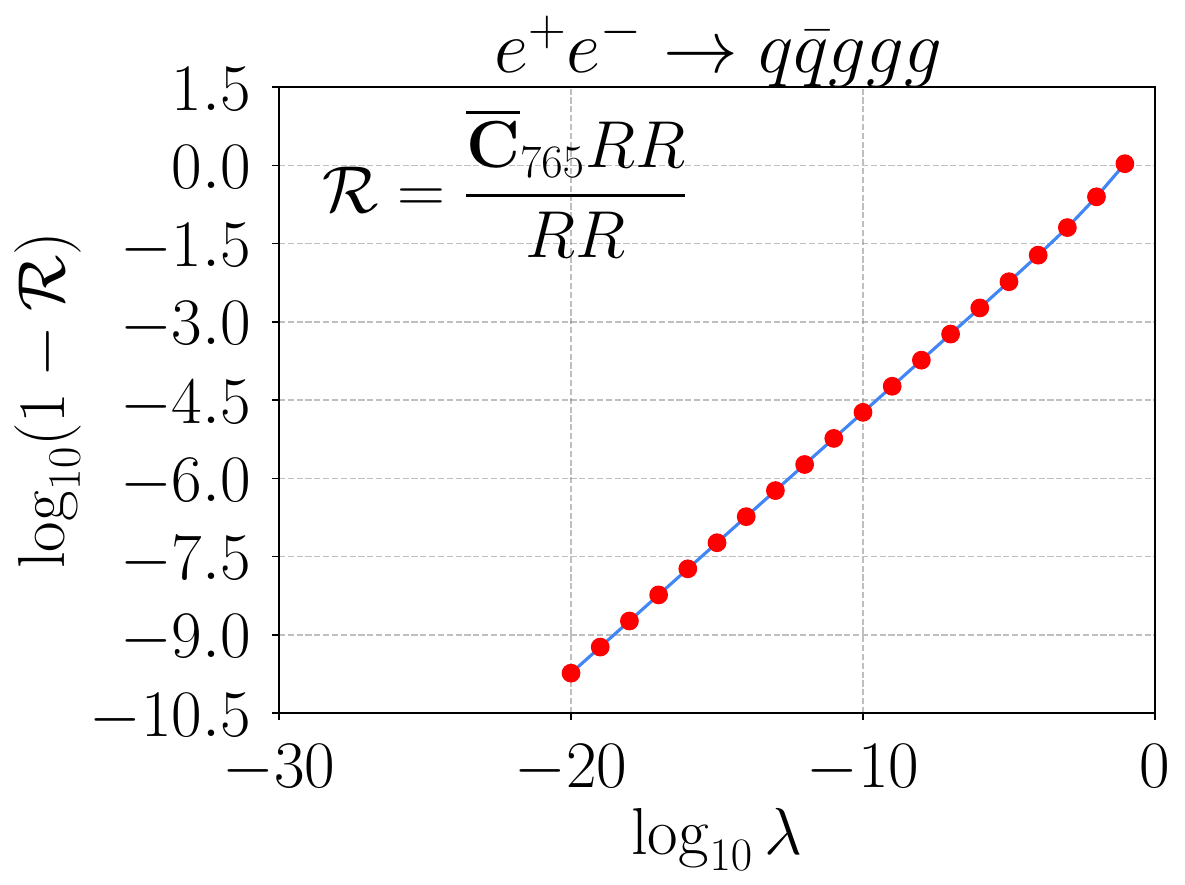}
\caption{{\label{fig:S76-C765sep}} $\LassSoft{76}{RR}$ and
$\LassColl{765}{}{RR}$ subtractions compared to SME in their limits,
respectively.}
\end{figure}

First we consider the individual subtraction terms for double-soft and
triple-collinear emissions that should be compared to the double-real
SME in the same limits. In case of correct definition and
implementation their ratios should tend to one as we go deeper and deeper
into the limit,
\begin{align}
  \frac{\LassSoft{76}{RR} }{RR}
  \xrightarrow{6\,,7\to 0} 1
  \,,\quad
  \frac{\LassColl{765}{}{RR} }{RR}
  \xrightarrow{5||6||7} 1
  \,.
\end{align}
The corresponding convergent set of ratios are depicted on
\fig{fig:S76-C765sep}.

\begin{figure}
\centering
\includegraphics[width=0.49\textwidth]{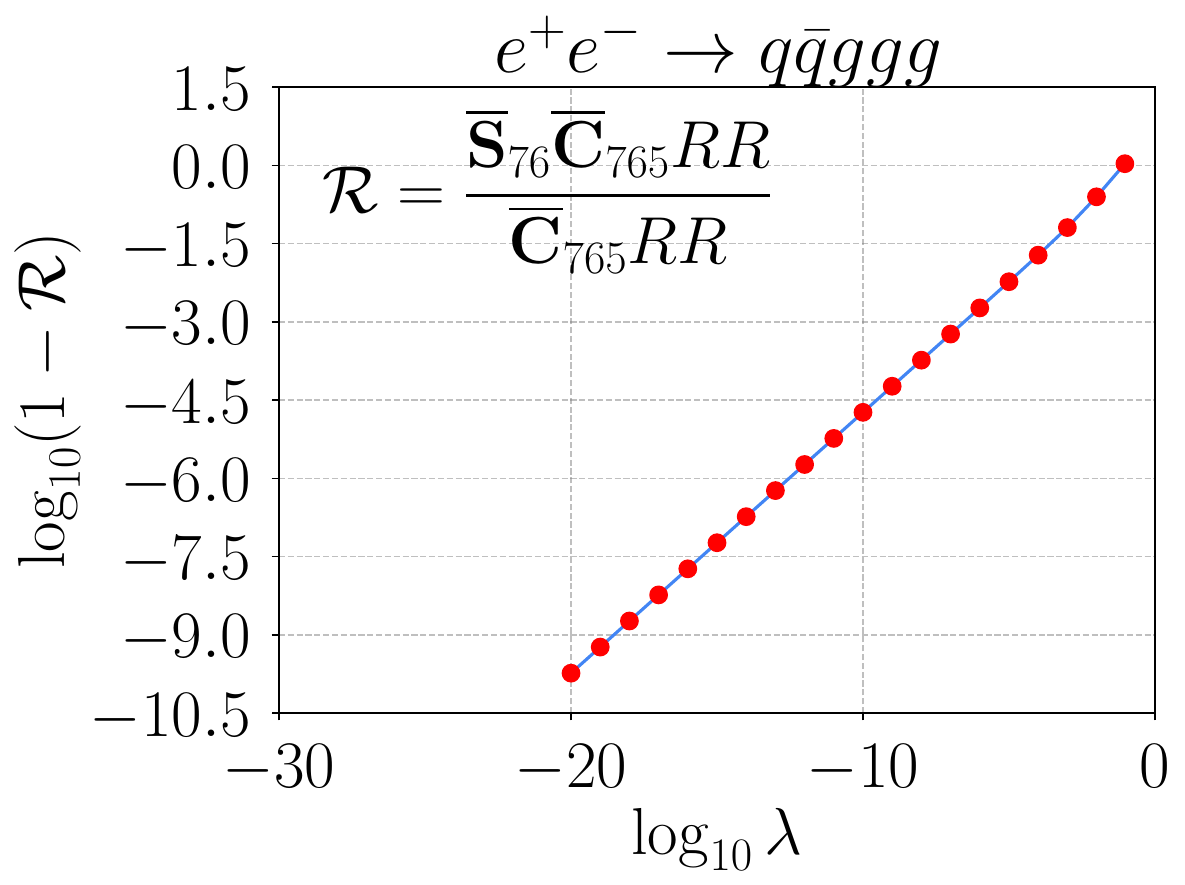}
~
\includegraphics[width=0.49\textwidth]{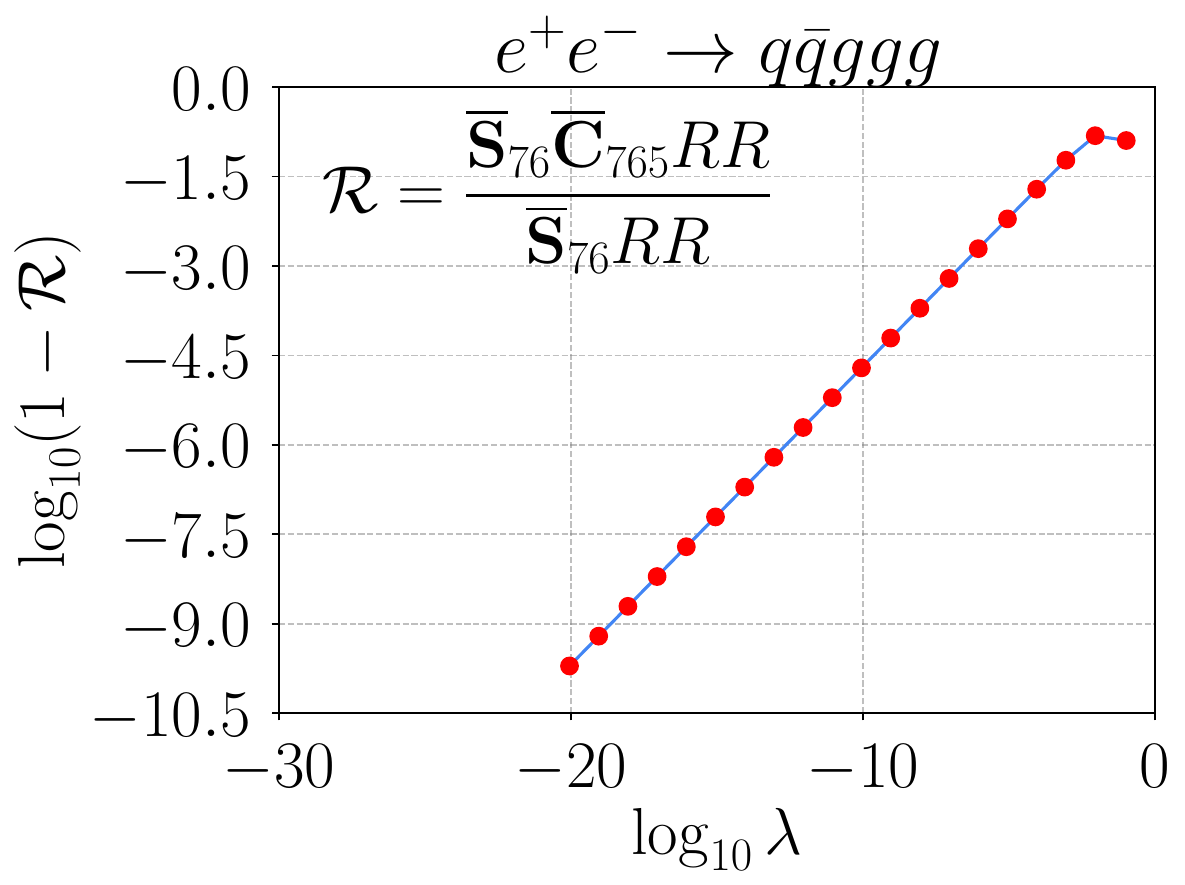}
\caption{{\label{fig:S76C765}} The $\LassSoft{76}{}\LassColl{765}{}{RR}$
compared to the corresponding triple-collinear (left) and double-soft
(right) term taking the appropriate limits.}
\end{figure}

There is an overlap between the double-soft and triple-collinear
subtractions. If the overlap term is well defined it should cancel with
the double-soft(triple-collinear) term in the
triple-collinear(double-soft) limits,
\begin{align}
  \frac{\LassSoft{76}{}\LassColl{765}{}{RR} }{\LassColl{765}{}{RR} }
  \xrightarrow{6\,,7\to 0} 1
  \,,\quad
  \frac{\LassSoft{76}{}\LassColl{765}{}{RR} }{\LassSoft{76}{RR} }
  \xrightarrow{5||6||7} 1
  \,.
\end{align}
The corresponding set of points can be seen on \fig{fig:S76C765}.

\begin{figure}
\centering
\includegraphics[width=0.49\textwidth]{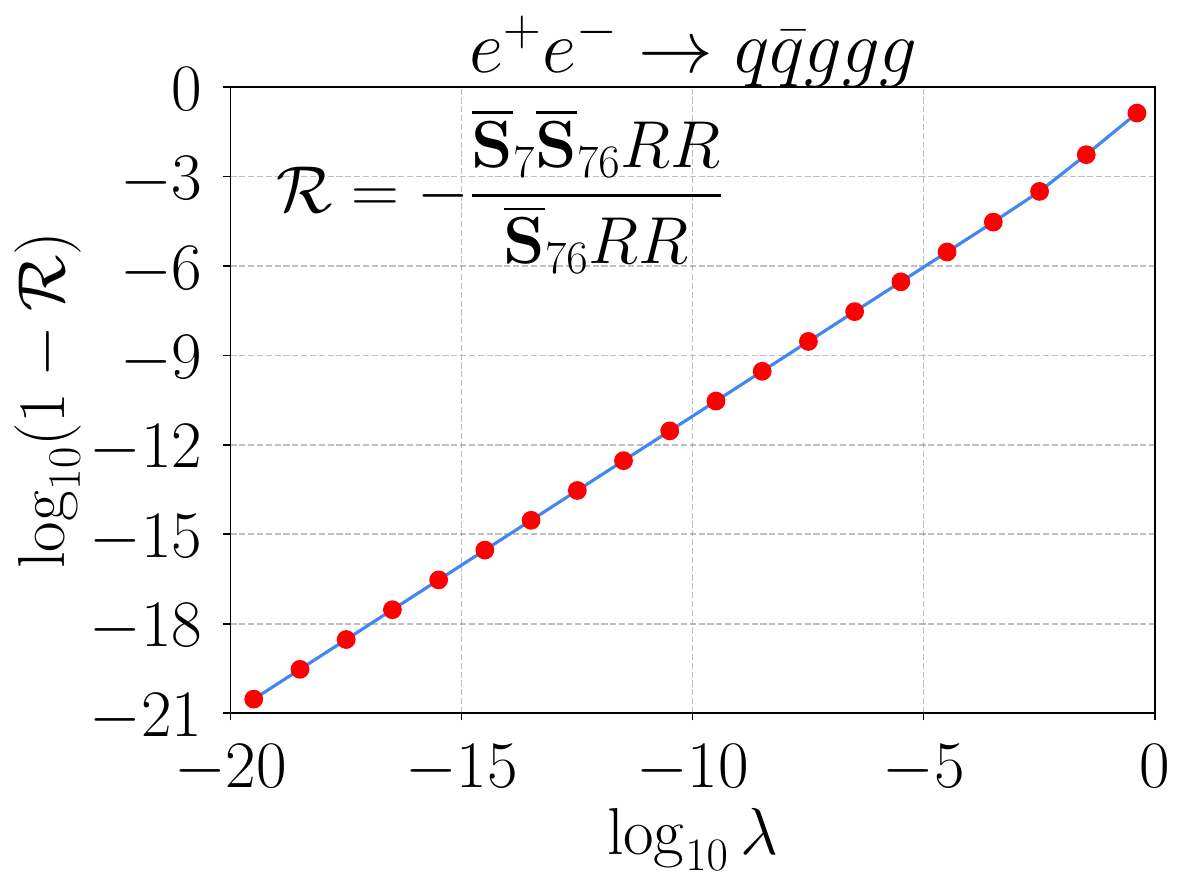}
~
\includegraphics[width=0.49\textwidth]{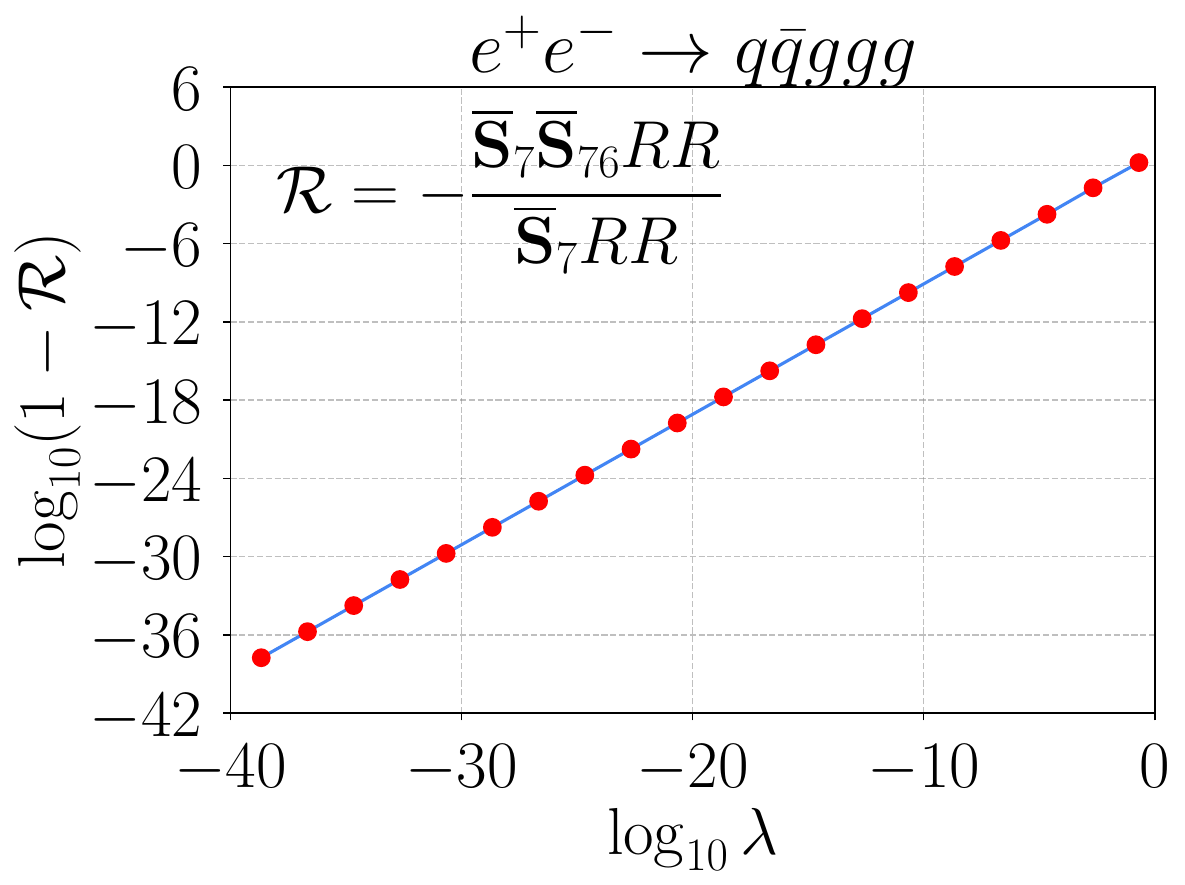}
\caption{{\label{fig:S7-S76}} Example convergence for a spurious
singularity cancellation involving soft limits.}
\end{figure}

To illustrate the cancellation of spurious singularities we consider a
pair of single- and double-soft subtraction terms. To avoid spurious
singularities a strongly-ordered soft term has to be defined with the
following conditions:
\begin{align}
  \frac{\LassSoft{7}{}\LassSoft{76}{RR} }{\LassSoft{76}{RR} }
  \xrightarrow{7\to 0} -1
  \,,\quad
  \frac{\LassSoft{7}{}\LassSoft{76}{RR} }{\LassSoft{7}{RR} }
  \xrightarrow{6\,,7\to 0} -1
  \,.
\end{align}
The corresponding plots can be seen on \fig{fig:S7-S76}. Notice that
the ratio approaches minus one in case of spurious singularity
cancellations. In our case the corresponding minus signs are defined
into the corresponding terms of $K^{(12)}$ which makes it easier in the
development phase to group subtraction terms in deep limits to observe
cancellations when happening in groups.

\begin{figure}
\centering
\includegraphics[width=0.49\textwidth]{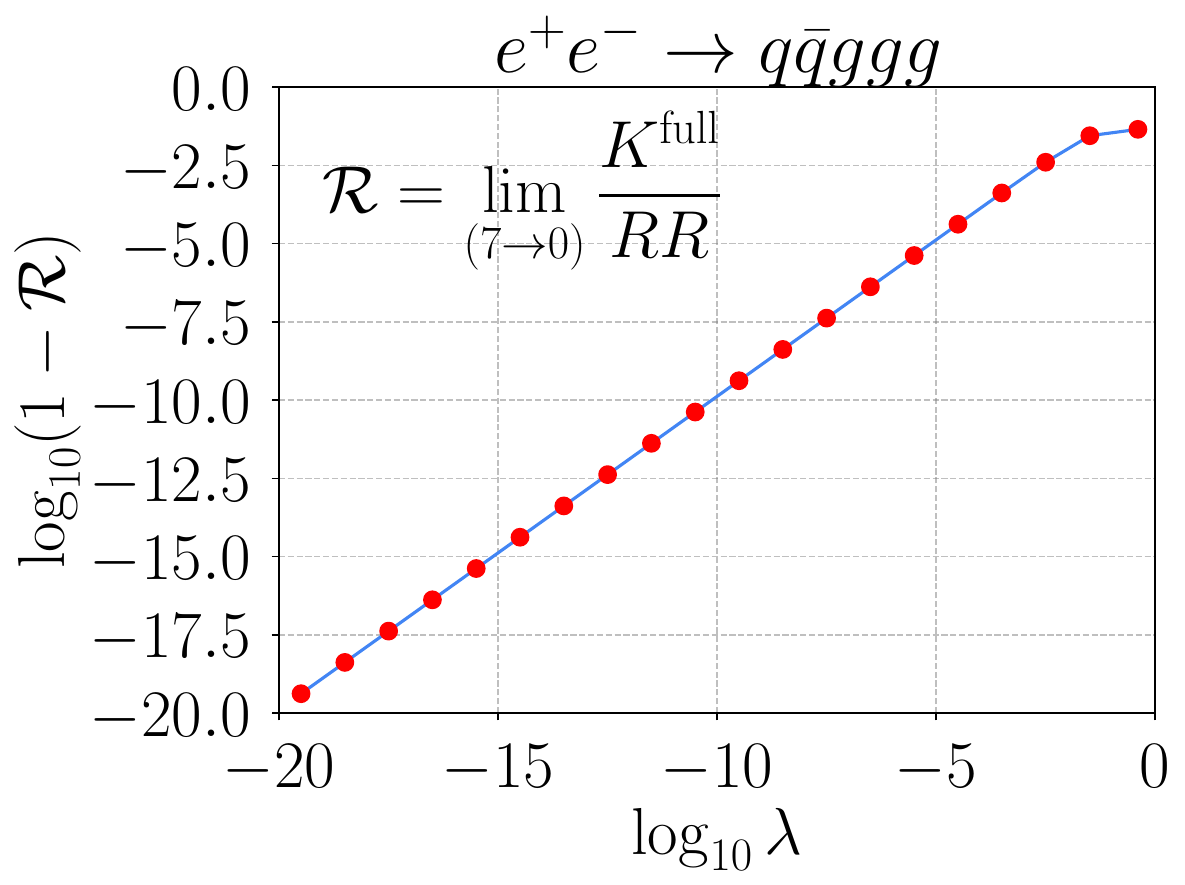}
~
\includegraphics[width=0.49\textwidth]{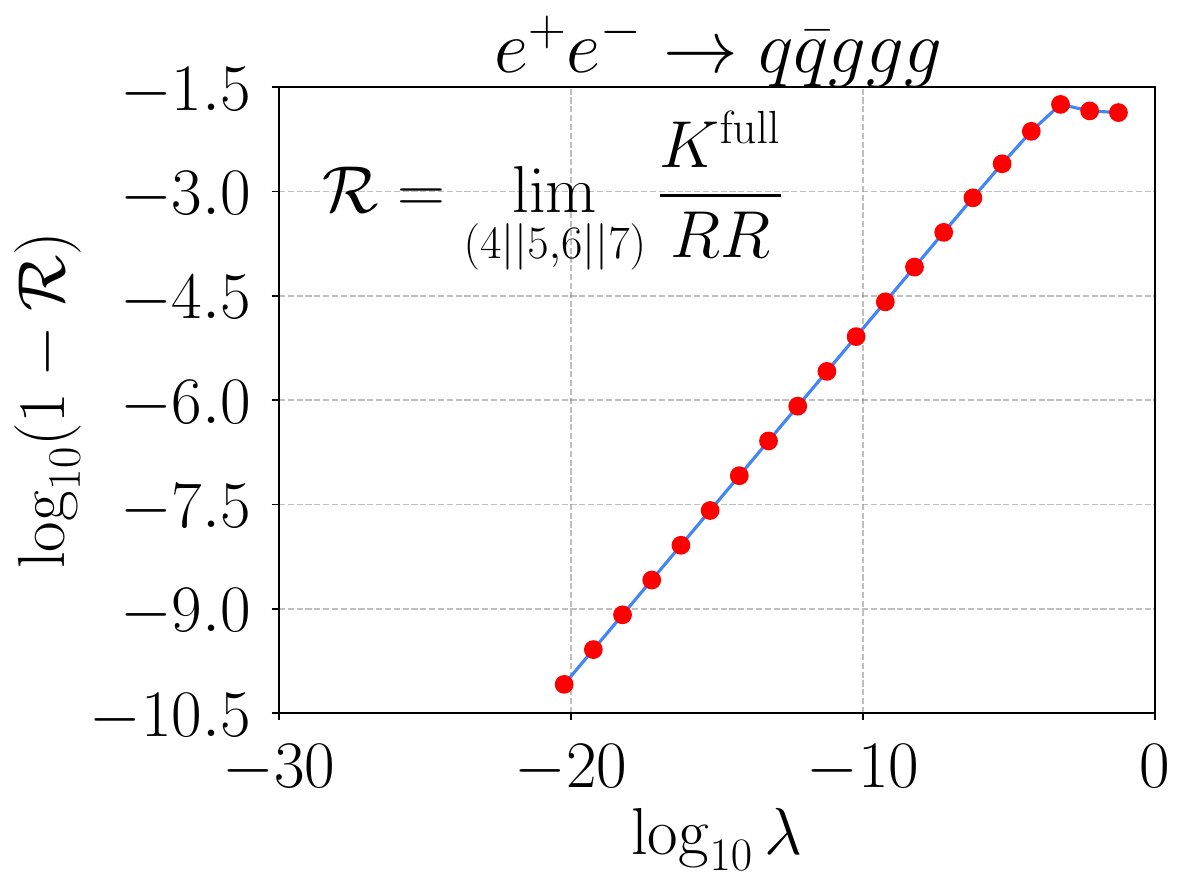}
\caption{{\label{fig:TotalSandC}} Comparison of the total set of
subtractions to the SME in two physical limits: single-soft (left)
and double-collinear (right).}
\end{figure}

Finally, we demonstrate the correctness of the complete set of
subtractions by considering a physical limit and comparing to the 
double-real SME. For illustrative purposes we consider a single-soft
and a double-collinear limit:
\begin{align}
  \frac{K^{(1)} + K^{(2)} + K^{(12)} }{RR }
  =
  \frac{K^{\rm full} }{RR }
  \xrightarrow{7\to 0}
  1
  \,,\quad
  \frac{K^{(1)} + K^{(2)} + K^{(12)} }{RR }
  =
  \frac{K^{\rm full} }{RR }
  \xrightarrow{4||5\,,6||7}
  1
  \,.
\end{align}
The corresponding plots can be seen on \fig{fig:TotalSandC}.

%}}} Numerical checks

\section{Conclusions}
%{{{
We presented the first numerical implementation of the Local Analytical
Subtraction Scheme by illustrating those numerical checks we conducted
in order to ensure our numerical implementation is capable of
regularizing all those kinematic singularities that can emerge in a
computation of a QCD cross section in electron-positron collisions
at NNLO accuracy. In our numerical studies we also investigated and
proved that subtractions are set up such that they do not only cancel
kinematic singularities of the radiation SMEs but also the spurious
singularities being inherently present at NNLO. To perform these
studies we created an arbitrary-precision numerical framework built
around the arbitrary-precision package written in \texttt{Fortran~90}
called \mpfun. Our results convinced us that subtractions are 
coded correctly and our next step towards the computation of a complete QCD
prediction at NNLO accuracy for distributions for three-jet production
in electron-positron annihilation is to start a numerical integration of
these terms with appropriate dynamics in a parton-level Monte-Carlo code.
%}}} 

\section{Acknowledgements}
%{{{
A.K. is supported by the UNKP-21-Bolyai+ New National Excellence Program of the Ministry for Innovation 
and Technology from the source of the National Research, Development and Innovation Fund. A.K. also kindly
acknowledges further financial support from the Bolyai Fellowship programme of the Hungarian Academy of Sciences.
G.B is supported by the Hellenic Foundation for Research and Innovation (H.F.R.I.) under the "2nd Call for H.F.R.I. Research Projects to support Faculty Members \& Researchers" (Project Number: 02674 HOCTools-II).
%}}}

\end{document}